\documentclass[12pt]{article}
\title{\bf Relative motion in spacetime}

\author { Philippe Droz-Vincent\\[2mm] {\sl LUTH}
 \footnote{Observatoire de Paris-Meudon, CNRS, Universit\'e Paris Diderot,
5  place Jules Janssen,   92195  Meudon, France   }}

\date{
   }

\newcommand  {\eeq}{\end{equation}}
\newcommand  {\beq}{\begin{equation} }
\newcommand  \half {  {1 \over 2} } 

\newcommand  {\ytil}{\widetilde y}

\newcommand {\ztil}{\widetilde z}
\newcommand {\hatx}{\widehat x} 

\newcommand {\hatr}{\widehat r}
\newcommand {\hatXi}{\widehat   \Xi }

\newcommand  {\noi}{\noindent}  

\newcommand  {\joka}{a^2}    
\newcommand  {\jokb}{b^2}
\newcommand  {\disp}{\displaystyle}
\newcommand  {\mun}{ {\mu  \nu } }

\newcommand  {\zer}{ {(0)} }
\newcommand  {\rel}{ {\rm rel} }
\newcommand  {\eff}{ {\rm eff} }

\newcommand {\vareps}{   \varepsilon  }
\newcommand {\Sig}{   \Sigma   }
\newcommand {\alp}{\alpha}
\newcommand {\albar}{\overline \alpha}

\newcommand{\del}{\delta}

\newcommand{\gam}{\gamma}
\newcommand{\lam}{\lambda} 
\newcommand{\Lam}{\Lambda} 
\newcommand{\ome}{\omega}
\newcommand{\Ome}{\Omega}
\newcommand{\Ebr}{\breve{E}}
\newcommand{\calx}{{\cal  X}}

\newcommand{\calp}{{\cal  P}}

\newcommand{\dron}{\partial}   

\newtheorem{prop}{Proposition}
 \newtheorem{theo}{Theorem}

\newcommand{\beprop}{\begin{prop}}
\newcommand{\betheo}{\begin{theo}}
\newcommand{\enprop}{\end{prop}}
\newcommand{\entheo}{\end{theo}}
\begin{document}
\maketitle \abstract{In Minkowski spacetime, we consider an isolated  system made of two pointlike bodies interacting at a distance in the nonradiative approximation.    Our framework is the covariant and  {\em a priori} Hamiltonian formalism of "predictive relativistic mechanics", founded on the equal-time condition.
The issue   of an equivalent one-body description is discussed.  We distinguish two different concepts:   on the one hand  an  almost purely kinematic {\em relative} particle,  
on the other  hand  an {\em  effective} particle which involves an explicit dynamical formulation;  several versions of the latter are possible.  Relative and effective particles have the same orbit, but may differ by 
their schedules.}

\bigskip

\section{Introduction}

The concept of  relative motion is basically  kinematic; it rests upon the  notion of  {\sl  radius-vector}  defined as the difference of the positions, it is viewed as the motion of a ficticious particle referred to as the "relative particle".
In nonrelativistic mechanics the two-body problem (concerning an isolated system) is easily reduced to a one-body problem concerning the relative particle (affected by the "reduced mass"). Of course,  in order to be equivalent  to  the binary motion,  
   the relative motion must be  considered {\em together with}  that of the center of mass. 

In the  framework  of special relativity  the situation becomes more complicated.  On the one hand extending to spacetime the construction of relative particle with help of a radius-vector is natural and  straightforward, provided we have a center of mass at our disposal.
On the other hand it  remains  to  be  checked  how  the result of this procedure  is  actually  equivalent to  the  binary motion. 

  In fact   the need of  considering an equivalent one-body motion first arose in the realm of  relativistic  {\em quantum} mechanics: in an early attempt by Itzykson  {\em et al.}  a  relativistic reduced mass
was suggested \cite{brezin}.      
 Todorov,  elaborating his  quasi-potential approach to the two-body problem      \cite{todmass}  ,  systematically invoked  an effective particle  supposed
 to satisfy a Klein-Gordon equation (with external field). 
The  {\em classical}  relativistic  version of  his work  \cite{tod} was   inspired from  Dirac's  constraint theory \cite{dira}; there  again, the task of  solving the equations of the binary system was  reduced to an
 effective one-body problem.  

\noi More recently    these topics were  considered in the framework of the "rest-frame instant form" of relativistic dynamics      \cite{lusanetal2},  but  due to a  different definition of center of mass,
the contact of our approach with that work would be far from straightforward.

For several authors the interest in relative motion was aiming  at  an approximate treatment of  the gravitational two-body problem in General Relativity  \cite{maheshTod} \cite{Dam} \cite{FizTod}, 
see also  \cite{Faruq}, 
but the scope of the present  paper is  strictly limited to Minkowski spacetime and special relativity. 
We have in mind the two-body conservative  dynamics of directly interacting point particles; the field carrying  interaction is supposed to be eliminated.  Following the  {a priori Hamiltonian} approach \cite{repor} \cite{annihp},
 we shall use the    manifestly covariant  formalism  of predictive relativistic mechanics \cite{droz} \cite{predico}, in the version where positions are defined by an equal-time condition;  
this point of view  offers several advantages:

\noi 1) it can  be re-formulated in terms of constraints, as shown in detail by  L. Lusanna \cite{luca}.

\noi 2) the relative orbit lays in a two-dimensional  (spacelike) plane. 

\noi 3) it allows  for  a covariant definition of center of mass   \cite{JMP1996}  which has the physical meaning  of Pryce's definition \cite{pryce}. 

Provided the existance of a linearly moving center of mass is assumed from the outset, the relative worldline always can be (in principle) constructed without difficulty.  In this procedure the input is a  two-body  motion, that is a {\em  pair} of worldlines,  and the output is a one-body  motion, that is a {\em single} worldline; at this stage we do not yet  consider  {\em systems}.

Somehow different is  the concept of "equivalent"  (or  "effective")  one-body  system. Given a binary system (defined by equations of motion derived from a Hamiltonian formalism) one aims at  finding  solutions,  and one hopes  to reduce this problem to the solving of a  single-body problem.                 One would expect that the relative particle defined on kinematic grounds is  apropriate   for this task.  
But investigating     
 under which circumstances  this  guess is correct  one runs into  severe conditions;  therefore we  shall  distinguish  "relative particle"  from   "equivalent one-body"  system,  keeping in mind however that,  for the sake of simplicity and efficiency,  the latter should   depart from the former as little as possible.   

  Section 2 is a summary of the covariant Hamiltonian formalism we  use for one-body and two-body systems.
Relative motion is analyzed in Section 3, and Section 4 is devoted to concluding remarks.

{\sl Terminology, notation}
Units  $c=\hbar =1$.  Greek indices $0,1,2,3$.  Latin indices $1,2,3$. 	Indicator $\vareps ^{0123} =1$.   When no confusion is possible tensor indices are omitted,  
and the contraction dot is employed also for tensors, so for instance 
$J \cdot P $ stands for $J ^{\alp \mu} P_\mu$.  Pointlike bodies are "particles"; particle labels are  $a,b$.

\section{One-body and two-body systems}

The motion of  a single {\em pointlike body} of  coordinates  $x^\alp$ is represented by its worldline. An inertial observer (of  unspecified mass) is characterized by a  timelike straight line of direction $u^\alp$.
Taking the origin of coordinates anywhere on the observer's   worldline,  the  {\sl orbit} of the body is, strictly speaking,  the set of  all the spacelike  four-vectors   $x_\perp ^\alp$,
where  $x_\perp  =   {\del _\perp}   x$ ,  using the projector
$$         {\del _\perp} ^\alp _ \beta =  {\del} ^\alp _ \beta     -   u^\alp   u_\beta $$
But naturally the orbit  can be trivially identified with the orthogonal projection of the body's worldline onto any three-plane orthogonal to  $u$.
 The {\sl  schedule } of the body consists in a relation between the points of the orbit and the observer's time.

\subsection{One body in a stationary  external potential}
Now consider a single {\em pointlike body} submitted to a {\em stationary } external potential  generated by a massive source at rest; this source is a distinguished inertial observer.
Using   the  Poisson brackets
\beq      (x^\alp , \   p_ \beta ) = \del ^\alp _ \beta   \label{PBone}   \eeq
we write the canonical equations of motion  in terms of  some evolution parameter $\tau$ ( most often  but not always, this parameter
 is affine; for instance it is {\em not}  when  the external field is a weak gravitational one)
\beq   {dx \over d \tau} =  (x,H) ,      \qquad   \             { dp \over d \tau } =  (p, H)                     \label{canon1}                    \eeq
The  Hamiltonian generator      
 \beq   H  =   \half  \ p ^2  + W            \label{defH}     \eeq
is a constant of motion numerically identified with  the  half-squared  mass.  $W$ is a scalar referred to as the "potential".

\noi  { Remark} $ \    $ In our  unconstrained  eight-dimensional phase space,    mass is not specified a priori, rather it
  appears as a constant of the motion; in this framework the dynamical system  encompasses all the possible numerical
 values of mass.

For our purpose it is convenient to separate space from time, with help of  the  projector 
$ {\del _\perp} $.
 The  assumption of stationarity  means that $W$ is a function of    only   $ \     x_\perp  ,   p_\perp , p_0 \         $. 
In any  frame  {\sl  adapted}  to $u$ we have that 
$$   (  x_\perp ,p_0 ) =     ( p_\perp , p_0 ) =  0    $$ 
so it is  clear that   the {\em spatial piece}  of the canonical equations of motion can be written as  well
\beq  {d \over d \tau}  x_\perp ^\alp = 
      (x_\perp ^\alp ,  \half  \ p_\perp ^2  + W )   \label{evolxperp}  \eeq
\beq  {d \over d \tau}  p_\perp ^\alp = 
      (p_\perp ^\alp ,  \half  \ p_\perp ^2  + W )   \label{evolpperp}    \eeq


\noi  Remark $ \   $ Since   $  ( x_\perp ^\alp , p_0 ) =0 $   we could replace $p_\perp ^2   $ by  $p^2$ in the
 r.h.s. of        (\ref{evolxperp})(\ref{evolpperp}). 

  The {\em timelike piece} of the canonical
 equations of motion  consists in
\beq   {dp_0 \over d\tau } =0              \label{evoltim1a}     \eeq              
\beq      {dx_0 \over d\tau } =   p_0   +   {\dron W  \over \dron  p_0 }   
 \label{evoltim1b}\eeq      
hence 
\beq     \    p_0 =   E = {\rm const.} \   \label{poE}     \eeq
which can be inserted into  system    (\ref{evolxperp})(\ref{evolpperp}).


\noi   Let us solve the spatial system      (\ref{evolxperp})(\ref{evolpperp})   above.  Since   $x_\perp$  and 
 $p_\perp$ have  vanishing Poisson brackets  with $p_0$, this quantity  can be treated as a constant, 
 say   $\    p_0  =  E \    $,  when computing the r.h.s. in 
  (\ref{evolxperp})(\ref{evolpperp}).          The task of integrating   these equations
 is similar to a nonrelativistic and three-dimensional problem. Beside initial data, its solution depends on $E$ as a parameter, say   
\beq  x _\perp =  \calx _\perp  (\tau , E )  ,  \qquad \quad
      p _\perp =  \calp _\perp  (\tau , E )    
 \label{solspace1}    \eeq
Finally we  integrate   eqn  (\ref{evoltim1b})
\beq   x^0 =  E \tau  +  \int  {\dron W \over  \dron p_0} \   d\tau  
\label{solx0}   \eeq
where the integrand is considered as a function of $\tau$ through  
(\ref{solspace1}) and $E$ behaves as  a parameter. We can summarize:

\beprop    In  any adapted frame,  the coordinate time of the body, namely $x^0$ is always the proper time (and also the coordinate time) 
{\em of the source},  whatever is the  mass of the source. 

\noi  In the simple  case where $ \dron W  / \dron  p_0 $ is identically zero~\footnote{for instance the toy model presented in \cite{IJTP} },
and using an  adapted frame,              
$x^0$ is proportional to the evolution parameter; in this case $E$ is necessarily positive. 
\enprop


\noi   Since  the potential is stationary,  energy $E$ (evaluated in the frame of the source) is conserved, see (\ref{poE}).
Since $2H = m^2$ we have
\beq   p _\perp ^2  + 2W    =  m^2  -  E^2        \label{m2E2}     \eeq
Setting   $p _\perp ^2 =  - {{\bf  p} }^2 $ we can introduce the nonrelativistic energy 
   $\disp  E_{\scriptscriptstyle  \rm  NR} = {  { {\bf  p} }^2  \over 2 m} -   {W  \over m}  $ which is obviously a constant of the motion, and
  we can rigorously write 
$$ E^2 =  m^ 2  +    2m    E_{\scriptscriptstyle \rm NR} $$
so the well-known formula $E =m  + E_{\scriptscriptstyle  \rm  NR} +  \cdots $    is obtained as  a development in
 powers of  $\disp   { E_{\scriptscriptstyle  \rm  NR} \over m}$ (and  the nonrelativistic limit  corresponds to  $E_{\scriptscriptstyle \rm  NR}  \ll  m$).

\noi
Rest energy simply   is  $ m$,   so we can define     {\sl unrest  energy} as   $  E -m$.

\noi  {Remark} $\     $
 Let a worldline be defined by
\beq   x^i = f^i (\tau ) ,  \quad  \quad    x^0  =  g (\tau )          \label{param}      \eeq
and let  ${\overline g} $ be   the inverse function of $g$, say  $\tau = {\overline g} (x^0) $.
Finally the  worldline is as well defined by 
  $$  x^i = f^i (   {\overline g} (x^0) )          $$  
Notice that if in    (\ref{param})  we   change only the  second  formula (in other words  we modify  $g(\tau )$,  leaving
$f^i$ unchanged)  then    we  keep    the same  orbit but  the schedule   is  modified, 
 and of course  the  worldline gets  changed.

\medskip
\subsection{Two-body isolated system}
The reader is referred to    \cite{annihp} \cite{IJTP} for details.
The Hamiltonian generators of motion are
$$  H_1 = \half p_1 ^2  +V_1 ,  \qquad    \                   H_2 = \half p_2 ^2  +V_2               $$
The interaction potentials $V_a$ 
 are functions on a 16-dimensional phase space and 
 the squared masses appear as first integrals,  say        $  2H_a  =  m_a ^2  $.
The canonical variables  $q_a^\alp,  p_b ^\beta$ are submitted to standard Poisson brackets 
$ \{ q^\alp _a ,  p_{b  \beta }  \}  =   \del _{ab}  \del  ^\alp _\beta  $.
 Due to a famous no-go theorem  \cite{Currie}, the canonical coordinates $q_a^\alp$
cannot coincide on the whole phase space with the physical positions $x_a^\alp$. 

\noi Notation$\     $  Collective and relative canonical variables are respectively
$$    P = p_1 + p_2,  \qquad   \        Q= \half (q_1  + q_2 )        $$
$$   y =    \half (p_1  - p_2 )     \qquad   \    z=q_1 - q_2               $$  
The system being isolated, the  interaction potentials $V_a$  are Poincar\'e invariant.  
Poincar\'e algebra is generated by  linear momentum $P$ and by angular momentum
$$  J =  q_1 \wedge  p_1    +       q_2 \wedge     p_2  =   Q  \wedge  P  +  z \wedge  y  $$

\noi  The relative  physical  position
$$      r  =  x_1  - x_2      $$
 should not be confused with $z$, {\em  except at  equal times}.

\noi Masses are $m_1,  m_2$.    It is convenient to set 
     \beq     \mu =  \half (m_1 ^2  +  m_2 ^2  )     , \qquad  \qquad      \nu =  \half (m_1 ^2   -  m_2 ^2  )    \label{defmunu}  \eeq 
Without loss of generality we assume  $m_1 \leq  m_2$.  Let  $m_0 = { m_1 m_2 /  ( m_1   +  m_2) }$
be the usual nonrelativistic  reduced mass; hence the inequalities
\beq    {1 \over 4}  m_1 ^2     \leq  m_0 ^2   <       m_1 ^2      \leq   \mu    \leq  m_2 ^2        \label{ineqs}   \eeq      
note that $m_1$ and  $m_0$ keep the same order of magnitude. 
The nonrelativistic  approximation is characterized by neglecting  $|\Lam |$ in front of {\em both} $m_1 ^2$  and  $m_2 ^2$. Post-Galilean developments 
can   be carried out  as  expansions in powers of  the dimensionless quantity $\disp    \Lam  /  m_0 ^2$.   

\medskip
The canonical equations of motion are
 \beq 
 {\dron  q_a  \over  \dron  \tau_b  } =   \{ q_a ,  \half    p_b ^2   + V_b  \} 
 \label{evolq}         \eeq
in terms of  individual evolution parameters    $\tau_1 ,  \tau_2 $ (generally {\em not} affine).
We define
$$  V  = \half ( V_1 + V_2 ) , \qquad  \qquad      U =  \half ( V_1 - V_2 )    $$  
Poincare invariance of   both  $V_1$  and     $ V_2  $ entails that   $P^\alp$  and  $J_ \mun $ are constants of the motion.

 \noi
The  coordinates of the     center of mass are
\beq  \Xi = Q +  ({y \cdot P \over P^2}) z   -    ({z \cdot P \over P^2}) y                \label{defXi}    \eeq
thus we have throughout phase space
\beq   \Xi \cdot P  =  Q \cdot  P                                                  \eeq  
Formula   (\ref{defXi})  admits two alternative and equivalent  expressions, namely
\beq  \Xi = {P \cdot p_1 \over  P^2} q_1  +   {P \cdot p_2 \over  P^2} q_2  +   {P \cdot  z \over  P^2}  y
\label{varvardefXi}  \eeq
which involves the {\em individual energies}
\beq  M_1 = {P \cdot p_1 \over  \sqrt{P^2}} ,  \qquad   \      M_2 = {P \cdot p_2 \over  \sqrt{P^2}}  ,    \label{defMa}   \eeq 
and  also
\beq  \Xi = {J \cdot P \over  P^2} +  \left(P \cdot Q \over P^2 \right) P     \label{vardefXi}  \eeq        
with  notation    $(J \cdot P )^\alp = J^{\alp \beta} P_\beta  $,         thus defining     
\beq    T  =  {P \cdot  Q  \over |P| }     \label{protimcdm}  \eeq
it turns out that  $T$ is the proper time of the center of mass~\footnote{ T has dimension of time, in contrast to the evolutions parameters $\tau_1 ,  \tau_2$.}

Fixing the total linear momentum (say
$P^\alp = k^\alp$  with   $k \cdot k  =  M^2$) defines the rest frame, where  $  \Xi ^0  =  T   +   {\rm   const.}\    $
   We shall always take the origin of coordinates  on the worldline of the center of mass,  therefore
\beq     \Xi ^0  =  T         \label{XioT}          \eeq
 
\noi  Notation    
   \beq       \Pi^\alp _\beta  =       \del^\alp _\beta      - { P^\alp   P_ \beta  /  {P \cdot P} },            \qquad   \              
              { \widehat  \del}^\alp _\beta   =        \del^\alp _\beta     - { k^\alp   k_ \beta  /  {k \cdot k} }         \label{projec}    \eeq

${\widetilde x}  =  {\Pi}  x , \qquad   \    {\widehat x}  =  {\widehat  \del}  x $,  etc.    

\medskip
 
   Consider the  equal-time description;  the equal-time manifold included in phase space is defined by $P \cdot z =0 $.

\noi   
In the rest frame:

 $z^0 = 0$  in other words   $q_1 ^0 = q_2 ^0$, thus 
  $x_1 ^0 = x_2 ^0$. On the other hand 
$$\Xi ^0 = Q^0 =  \half (q_1 ^0 + q_2 ^0 )    =     \half (x_1 ^0 + x_2 ^0 )    $$
 hence          
   $      x_1 ^0 = x_2 ^0 =  T      $.

\noi     In general we have 
\beq    y \cdot P = \nu - 2U                               \label{yP}    
   \eeq  
 so finally the {\em equal-time description}  takes on this form
\beq  \hatx _1 =  {\widehat \Xi }  + 
( {2U  - \nu  \over M^2 } + \half )   r  \label{x1space}   \eeq
\beq  \hatx _2 =  {\widehat \Xi }  + 
( {2U - \nu  \over M^2 } - \half )   r     \label{x2space}   \eeq
\beq   x_1 ^0  =  T,  \qquad  \     x_2 ^0  = T        \label{x12time}         \eeq 

\medskip
\noi 
{\sl Definition}$\qquad \      $The  {\sl orbit of a two-body motion} is the set of values taken by  $\hatr$ in  the  three-dimensional vector space orthogonal to $ k$.
 
Owing to the constancy of angular momentum, it is a plane curve included in the two-dimensional vector space  orthogonal  
 to linear momentum  $P^\alp$ and to the   Pauli-Lubanski vector
\beq       L_\alp  =     \vareps    _{\alp \mu  \nu  \rho  }  P^\mu  \ztil^\nu   \ytil^\rho          \eeq

\subsection{Unipotential models}
Several simplifications arise  when $U $ is identically zero on the whole phase space, in other words $V_1 = V_2 = V$.   In this case
\beq   V = V (\ztil ^2 , \ytil ^2 ,   \ztil  \cdot     \ytil ,       P^2  ,    y \cdot   P  )            \label  {unipot}    \eeq
and it automatically turns out that       $  y \cdot   P$ is  a  constant  of the motion.
\
\noi  This particular model, referred to as {\sl unipotential}, is still general enough for describing a lot of physical situations. In this case 
it is convenient  to  set
\beq    \lam = \tau_1 + \tau_2                            \eeq
so the equations of motion for  $ \ztil ,  \ytil $  can be written as  \cite{cras}
  \beq 
 {d \ztil  \over  d \lam  } =   \{ \ztil ,  \half    \ytil ^2   + V  \} 
 \label{evolztil}         \eeq
 \beq
 {d \ytil  \over  d \lam  } =   \{ \ytil ,  \half    \ytil ^2   + V  \} 
                    \label{evolytil}              \eeq
where the brackets can be computed as  functions of   $ \ztil ^\mu , \ytil ^\nu ,  $   and  of  the first integrals  $P^2 ,   \     y \cdot P$.
 Once  $P^2$ and $y \cdot P $ have been fixed,
the evolution of the  spatial internal variables is thus given by a system of six 
first-order differential equations, to solve for six unknown functions; this problem has the structure of a {\em nonrelativistic}
  problem  for one body in  three dimensions. 

\noi  Remark:   In the evolution equations above,  $P^2$ and  $y\cdot  P$ have vanishing Poisson brackets  with  $ \ytil^2  + 2V $ 
(remember V doesnot depend on     $z\cdot  P$), thus  they behave like constants when computing  the right-hand sides. 
The four-vectors    $\ztil$ and     $\ytil$ remain within the 2-plane  orthogonal to $k$ and to the (conserved)
 Pauli-Lubanski vector. 
Note this  constant of the motion
\beq     j^2  =  \ztil^2     \ytil ^2     -   ( \ztil \cdot   \ytil )^2        \label{defj2}       \eeq  

\medskip
   Let  a  solution   of  (\ref{evolztil})(\ref{evolytil})        be of the form   
    \beq    \ztil =  \zeta (\lam , P^2 ) ,  
 \qquad  \  
        \ytil =  \eta (\lam , P^2 )   
        \label{solspace2}              \eeq  


 For any unipotential model it turns out that   $N= \ytil ^2   + 2V  $ is a constant of the motion;
 fixing $k^\alp$ and the individual masses 
 amounts to fix  its numerical value, say $N= - \Lam$,   where 
\beq   \Lam = {M^2  \over 4}   +{\nu ^2  \over M^2}  - \mu     \label{defLam}          \eeq
is intimately related to  properties of   relative motion~\footnote{This quantity appeared, denoted as  $b^2$ in the work of Todorov \cite{todmass} }.
The  inverse  formula
\beq    M^2  =  2 (\mu + \Lam)  +  2  \sqrt {(\mu + \Lam)^2  - \nu ^2}   \label{MdeLam}      \eeq
entails the development
\beq  M =  m_1  +  m_2   +   {m_0 \over 2}  \         {\Lam   \over m_0 ^2}   +   O( ({\Lam   \over m_0 ^2})^2 )    \label{devlopM}    \eeq    
hence     $\    M \simeq    m_1  +  m_2  , \      M_a \simeq  m_a  \      $  in  the  nonrelativistic limit. 

 
\medskip

As proved in  \cite{IJTP} ,  in order to achieve a parametric description of the worldlines in terms of $\lam$   we have to complete  
(\ref{solspace2})  with  
\beq   T =   ({M \over 4} - {\nu ^2 \over M^3}) \lam +  {1 \over M} \int F d\lam   - { \nu \over M^ 3 } \int G d\lam 
        \label{Tcdm} \eeq
where 
\beq  F (\lam, M^2 , \nu )  =   \{ Q \cdot P  ,  V \},    \qquad  \           G  (\lam, M^2 , \nu )  =   \{ z \cdot P  ,  V \}           \label{defFG}    \eeq          
Remark  $\   $   if  $V$ doesnot depend on  $y \cdot P$ then   $G$ identically vanishes.


\medskip       
   \section{Relative  Motion}               
Before proceeding further let us recall a few features of relative motion in  {\em nonrelativistic  mechanics}, for an
{\em isolated} system of two point particles:

\noi a) Relative motion is essentially      
 the time evolution of the "radius-vector" joining the constituent bodies of the system.
Obviously the   values  of this  radius-vector  are not points of  an affine space but  elements of a three-dimensional  vector space;
 however it is natural to take the center of mass as origin of space, which leads to construct the positions of the relative 
 particle moving "around it".

\noi b) The motion of the center of mass gets completely separated from relative motion.

\noi c)  Equivalence:   knowledge of the binary motion  determines both  the relative motion and the center-of-mass motion, 
 {\em and vice versa}.
  
\noi  d) The mass of this ficticious particle, referred to as the {\em reduced mass}, is unambiguously determined from the individual
 masses of the constituent of the system,  say
$ m_0 = m_1 m_2 / (m_1 + m_2) $,  through the  separation of   center of mass from the relative degrees of freedom.

\noi  e) In the limit where the mass ratio   $m_1 / m_2  \rightarrow 0$,  the light body comes to coincide with the relative particle.

\noi
f)  Relative motion is  viewed as submitted to external forces, derived from an external potential.
With the convention just made in  (a) above,  this potential can be seen as created by a ficticious source located at the center of mass.

\noi   g)  Hamiltonian formulation holds as well for the binary system  as  for the relative particle, and the correspondance is ensured
 by well-known formulas.


\medskip
Remark $\    $Note that the  points (a)(b)(c) 
 are strictly  kinematic, in this sense  that   they do not  refer    to  a   canonical   formulation.
In contrast   the reduced mass invoked in  (d)  rests on the separation of    internal  {\em vs}   external degrees of freedom, in the Hamiltonian framework.
Finally (f)(g) explicitly  refer to the  Hamiltonian formalism.

Remark. $ \      $ In nonrelativistic mechanics the concept of  center of mass (once the masses of the bodies have been fixed) is of pure  kinematics, we could almost say pure geometry. 
In contrast, in special relativity  one runs into the problem of spacetime foliation; the way out is clear provided the binary system is isolated:  linear and   angular momenta are conserved;
 the former provides a preferred foliation and the latter can be combined with the  former  in  formula     (\ref{vardefXi})   which yields the center-of-mass coordinates.   
The principles  of this procedure have been discussed many years ago by Pryce \cite{pryce}  and by Moeller \cite{moell}.
But  some dynamics is already implicitly  involved here.


\medskip
Remark. $ \    $  In classical mechanics   the relative particle  is definitely  the most simple effective one-body 
 system equivalent to the binary system.

$$ \     $$

Now we return to {\em special relativity}, with canonical equations of motion   (\ref{evolq}). 
We aim at giving a clear definition of relative motion in spacetime. We cannot expect that all the features of the classical theory can be generalized easily.  
 So, as a first step, we start with a purely kinematic construction of the relative motion.

\subsection{Kinematics, relative particle.}
In this subsection  we provisionally forget   our       canonical equations of motion   (\ref{evolq}).   
      A motion of two pointlike (structureless) bodies is essentially a pair of worldlines.
At  this stage we need not specify the dynamics, except for     

\noi  {\bf Assumption A1} $\    ${\em There exists   
 a  center of mass, with definite mass $M$ and  constant  momentum $k^\alp$ (such that  $k \cdot k = M^2$)
moving along a (timelike) straight line}.

\noi
 Let $\Xi^\alp$ be the center-of-mass coordinates;  we can write
\beq  \Xi^\alp = \Xi_\zer ^\alp +  T   \   {k^\alp \over M}      \label{3wardefXi}       \eeq
$\Xi_\zer$ being an origin taken on the center-of-mass  worldline; automatically  $T$ is the center-of-mass proper time.
The rest frame of the system is determined by the direction of $k^\alp$. 
Given  $k^\alp$ we have a distinguished slicing of spacetime by three-planes orthogonal to it.
This allows for an {\sl equal-time description of motion},  such that   $x_1$ and $x_2$, hence also  $r =x_1 -x_2$,  are functions
 of the center-of-mass time $T$, whereas   
\beq    x_1 ^0  =  x_2^0  =   T           \label{xa0T}                  \eeq
(this formula, 
  stems from equal-time description, irrespective of how interaction is modelled). 

\noi
 The four-vector  $ r =  x_1 - x_2$ will be called  be the {\em radius-vector}; it  is  spacelike and  has the usual physical meaning when it is orthogonal to linear momentum $k$.
By itself it defines no point in the (affine!) Minkowski space, but in the spirit of   (a)  listed above,  
\noi {\sl Relative motion can be though of as that of a ficticious particle with  position $x^\alp  _\rel$ defined by attaching the radius-vector
 to the center of mass}, in other words   the relative particle is unambiguously defined  by 
\beq    x_\rel =  \Xi + \hatr                   \label{defeob}            \eeq

\noi
The motion of the bodies will be described by  the sequence of the couples $x_1 , x_2$ taken on the worldlines and {\em simultaneous} with 
respect to  the slicing of  spacetime determined by $k^\alp$.  

 \noi Since we consider only simultaneous positions, $r$ is orthogonal to $k$, and in the  rest  frame we have $r_0 =0$ and
  $x_a^0$ as in  (\ref{x12time})


\noi Separating time from space in (\ref{defeob}) we get
\beq      x^0 _\rel  =   \Xi ^0    \label{28bis}      \eeq  
and    $             \hatx _\rel =    {\widehat  \Xi  } +  \hatr      $   which becomes, by a  choice of the origin such that
$\Xi ^\alp _\zer = 0$,
\beq      x^i _\rel   =  r^i                                        \label{28ter}      \eeq                 

\noi  But  $\Xi ^0 $ is the center-of-mass time, thus in view of (\ref{varvardefXi}) equation  (\ref{28bis})   simply     means
        \beq     x^0 _\rel  =  T    \label{tcdmT}   \eeq

\noi Now the motion of the relative particle  is   defined by (\ref{28ter})(\ref{tcdmT}).  Projection of the relative worldline onto the three-dimensional plane
 $T=0$ may be trivially  identified with the relative orbit.

  So far  we have constructed  the relative  particle  as  undergoing  a  ficticious one-body motion, characterized by its worldline; 
now conversely, given this worldline    and that of the center of mass,   is it possible to 
 {\em  reconstruct} the initial couple of worldlines, say   $x_1 (\tau_1 ) , x_2 (\tau_2 ) $ ?
    Only {\em  if the answer is yes}, the relative-particle motion  can be considered as 
{\em equivalent} to that of the binary object.

Notice also that, although  the relative particle undergoes a  perfectly well-defined motion,   kinematics alone provide no hint  about its {\em energy} or its {\em mass}; these quantities  could  be  eventually  exhibited  if we were able of  setting   a  {\em one}-body canonical formalism  that pure kinematics ignore.

In fact we have elaborated  in  kinematic terms a  correspondance    {\em from  motion to motion}. 

\medskip
In order to address the issue of equivalence  (which  regards  the possibility  of  {\em reconstruction} of the worldlines)  let us now be  more specific and assume 

\noi  {\bf Assumption A2} $\    ${\em Considered at equal times, the center of mass  is aligned with $x_1$ and $x_2$  and 
lies between them.}  

\noi This statement may be expressed as follows 
\beq   \hatx_1  -  \hatXi = b  \hatr , \qquad  \           \hatx_2  -  \hatXi = -a  \hatr         \label{A2}         \eeq
where  $a$ and  $b$ are positive {\em functions of}  $T$  such that  $a +b =1$. 

\noi It follows immediately that 
\beq  \hatXi  =   a  \hatx _1  +   b  \hatx _2       \label{Prycelike}    \eeq

In order to re-construct the worldlines of the binary motion we are provided with    (\ref{A2})  and  the equal-time condition (\ref{xa0T}).
The only truble is about the space part: though $\hat\Xi$ and  $\hatr$ are  given  inputs, solving (\ref{A2}) for $\hatx_1$ and $\hatx _2$  still  requires knowledge
 of the coefficients $a, b$.  As obvious in (\ref{Prycelike}) these coefficients are the ingredients of  $\hatXi$, but starting from $\Xi$ and  $x_\rel$,  it is generally  not possible to revert back to $a$ and $b$. 

\noi  Remark $ \    $  In the most generic case,  knowing  also  the individual masses $m_1 , m_2$  would not help:
 $a - b$  is   some dynamical variable of the binary motion, not always a constant~\footnote{In the special  case where  $a-b$ is a 
constant of the motion,  an alternative problem  would consist in giving 
 $a$ and $b$ instead of the individual masses; this problem is already and trivially solved by ({A2}); note that  requiring  constancy of  $a-b$  is  only  a little more general than   assuming a  unipotential model.}
 Further information about   $a$ and $b$  requires specifying how is the center of mass defined within the underlying dynamics, 
 which goes beyond the  kinematic  approach proposed in the present  subsection,  so 

\noi{\em In purely kinematic terms, with center of mass and relative particle as the only data,  re-constructing  the  worldlines of binary motion  is  
generally {\em not}  possible}.

For an illustration, let us now focus on  a two-body system  described by
  the  {\em a priori } Hamiltonian formalism of predictive relativistic mechanics, 
with help of the equal-time prescription;  for the moment $U$ is possibly nonzero.
Remember              
(\ref{yP}).  Then   formula (\ref{defXi}) tells that  
   {\em at equal times}    we can write 
 $ \disp  \Xi = \half (q_1 + q_2 )+  ({y \cdot P \over P^2}) z  $.
 But then  $q_a = x_a$,  and  $z=r$. Finally we get    
\beq  x_1 = \Xi  +  (\half -  {y \cdot P \over M^2 } ) \   r    \label{indiv1}  \eeq 
\beq   x_2 = \Xi  -  (\half +  {y \cdot P \over M^2 } ) \   r    \label{indiv2} \eeq
implying 
\beq \hatx_1 = \hatXi  +  (\half -  {y \cdot P \over M^2 } ) \     \hatr    \label{hatindiv1}  \eeq 
\beq \hatx_2 = \hatXi  -  (\half +  {y \cdot P \over M^2 } ) \      \hatr   \label{hatindiv2} \eeq
that is an example of formulas  (\ref{A2}), where  
    \beq             {y \cdot P  \over P^2}   = { a-b  \over 2}       \label{yPab}   \eeq

\noi
In these formulas $r^ \alp$ is given by the worldline of the "relative particle".  Indeed this worldline can be parametrized with
 help of the center-of-mass time, say  (\ref{28ter})(\ref{tcdmT})   where  $r^i$  takes on the form      $  r^ i =  \phi^ i (T)    $, 
 more general than  $\zeta ^i $ , since here $U$ is not yet assumed to be zero.

In contrast             the quantity   $y \cdot P$  is  {\em generally not} fully  determined by    knowing  relative motion and center-of-mass motion,  
because of the contribution of $U$ to it. Intuitively we could put it that way:  $U$ carries dynamical information which goes beyond simple kinematics.


 Naturally this difficulty disapears  if we decide to {\sl  focus on  unipotential models}, as we do in the sequel;
  indeed  now  we can write 
$$ a  =  
 {M_1 \over M} ,   \qquad       \        b  = 
 {M_2 \over M}      $$
 It follows that  
     $$    a-b  =  2  {y \cdot P  / M^2}  =  2 \nu  /  M^ 2 $$
 which is a constant and depends only on  the  masses
 $M, m_1 , m_2$ (note that  $a \leq  b$). So  we can state

\betheo  For {\em unipotential models},  knowing the relative motion,   its   worldline and the mass of the 
center of mass,  plus the squared-mass difference~\footnote{This  is a little  more general than assuming that
we  know   {\em both} individual masses.} ,  amounts to knkow 
 both worldlines of the interacting bodies.
\entheo
In the  present case   we can say that the one-body  motion   of the relative particle   is   equivalent   to  the  binary  motion.

\medskip

We stressed in  \cite{IJTP} 
  the importance of having {\em  both} individual energies positive, {\em as we assume henceforth},  which amounts  to  demand
\beq   m_1 ^2  +  \Lam    >  0          \label{new48}      \eeq
Under this condition  we proved
\beprop   If  we can neglect  $ \sqrt {|\Lam |} $ in front of  $m_2 $,  then  we have that, {\em at equal times}  $\Xi$ and  $x_2$  coincide  in the limit  $m_1 /  m_2   \rightarrow   0$.
\enprop
This statement  was labelled as "theorem 2" in \cite{IJTP}.   Since  $\Xi$ and  $x_2$ are coordinates of points  describing timelike curves,  their coincidence at equal times 
implies coincidence of the  worldlines; this  amounts  to say that, taking  $m_1 /  m_2  $  to zero,      the worldline of the center of mass  coincides with  that of the  heavy particle.
We define $\disp  \gam =  {m_1 \over  m_2} $ and      $ \vareps  =  \gam ^2$. 

\noi     In fact the  assumption  made about  $\Lam$  
   is superfluous for   negative  $\Lam $,  owing to       (\ref{new48})  which  requires     $|\Lam |  <   m_1^2  =   \vareps  m_2 ^2$ .
  So    in this case   we can write    
$$ \Lam =  a  m_1 ^2 =  a  \vareps  m_2 ^2$$   
where   $|a|   < 1$. 
Although $\Lam   \rightarrow 0$  with  $\vareps$, a glance at (\ref{ineqs}) shows that $ \disp {|\Lam |  \over m_0 ^2}$   remains $> |a|$, so 
 the situation is clearly distinct  from a nonrelativistic regime.

\noi   In contrast   for  positive $\Lam$  neglecting   $ \sqrt {|\Lam |} $ in front of  $m_2 $  is  essential.  If we fix {\em a priori}  $\Lam$  independent  from $\gam$, 
the recoil of $x_2$ generally does not vanish when  $\gam \rightarrow 0$.
In contrast  taking   $\Lam = O(\gam ^p) $, for some positive power of   $\gam$, entails  $M   \rightarrow   m_2,  \          x_2    \rightarrow   \Xi $, etc (see §5  of   \cite{IJTP}).

 \noi   Vanishing  $\Lam$ is a trivial case. 

\medskip
 With help of the above  proposition  (and  under the same assumption)  
 let us  derive the following 

\beprop           
 Neglecting   $ \sqrt {|\Lam |} $ in front of  $m_2 $, we have that
  $  x_1  \rightarrow   x_\rel$  when   $m_1 /  m_2   \rightarrow   0$, in other words the light body  
comes to coincide with the relative particle.
\enprop  

\noi  Proof $\     $  We use the {\em equal-time} description. On the one hand definition (\ref{defeob}) of $x_\rel$    yields  
$$  \hatx _\rel =  {\widehat \Xi }  - \hatx _2  +  \hatx_1       $$
but, 
Proposition 2  entails  
${\widehat \Xi }  - \hatx _2    \rightarrow  0 $
hence        $ \hatx_1   \rightarrow   \hatx _\rel    $.
 On the other hand,  working at equal times, 
formula (\ref{xa0T}) holds true;  we also have have   (\ref{tcdmT}), therefore 
$  x_1 ^0   =         x_\rel ^0 =  T$.    Finally  $ x_1   \rightarrow      x_\rel $.                     []

\medskip
\noi  So properties  c)  and  e) of the list in 2.3  are  extended to the relativistic realm.
In contrast extension of   d)  f) g)  remain problematic   for,  in the kinematic context,  there is no indication
as to know whether  the {\em relative motion also} can be derived from a Hamiltonian
 of its own.  This question  leads  us  to dynamical considerations.   

Before that we turn to dynamics,  let us summarize the equal-time description of  the  relative particle. 
On the one hand we have, for the spatial relative variables,  the evolution equations (\ref{evolztil})(\ref{evolytil}).
On the other hand we have    (\ref{tcdmT}),  but now   the unipotential assumption  entails that   $T$ is given by    (\ref{Tcdm}).
 Defining   $\check{E}$
\beq  \Ebr = {M\over 4} -  {\nu ^ 2  \over  M^ 3}      =           { (P \cdot p_ 1 )( P \cdot p_2 )    \over   M^ 3}    =  {M_1  M_2  \over  M}     \label{defEbr}     \eeq 
(note that $\Ebr$  cannot be negative) we re-write   (\ref{Tcdm}) as follows
\beq   T  =T (\lam )  =   \Ebr  \lam + {1 \over M} \int F  d \lam 
                          - {\nu \over M^3 }   \int G d \lam      \label{Tdelam}    \eeq
According to   (36)   and  using   $r^i = z^i$ (equal times),  we have  
\beq          x^i  _\rel = \zeta ^i  (\lam , M^2 ,  y \cdot  P )           \label{eobspace}     \eeq
where  $\zeta ^i $ are  three  functions defined in    (\ref{solspace2}) as solutions of  (\ref{evolztil})(\ref{evolytil}).
 Moreover  by    (\ref{tcdmT})
\beq         x^0 _\rel=  T(\lam )         \label{eobtim}       \eeq  
In the context of our present assumptions, equations  (\ref{eobspace})(\ref {eobtim})    characterize the equal-time 
description of the relative particle.

Although  we considered a Hamiltonian model of the  binary object, the  motion of  our   relative particle {\em  is not derived} from any 
 canonical formalism.  At  this stage  there is no  indication about  its  mass~\footnote{In the absence of a canonical formulation, enforcing  $m=m_0$
 would not be  here justified  as much as it is in the  nonrelativistic domain.}, in other words the  concept of a reduced mass is still lacking.

\subsection{Dynamics, effective particle.}
In the previous subsection devoted to kinematics we were interested in  two-body  and one-body {\em  motions}.
 Here we come to dynamical {\em systems}, keeping clear in mind that  a system essentially is a collection  of possible motions.   

\noi  Now we consider an alternative approach to the problem of reducing binary motion to a one-body problem.

\noi In a more general setting, what  we search  now is a ficticious one-body  {\em   dynamical system} of which the solution
 gives knowledge of  the binary motion; in the most simple and the most reasonable manner. We mean a canonical formulation  
 of the kind displayed in Section 2.1.

More precisely, for any  fixed  value  $k^\alp$ of linear momentum, and setting $k \cdot k = M^2$,  we look for a 
one-body  Hamiltonian $H$,  depending  on    $k^\alp$ as a parameter,
 such  that it  generates the motion of a {\em ficticious }  particle referred to as {\sl effective} .
 To be more specific   we  require that  the external field (involved 
in the motion of the ficticious particle) is {\em stationary}, and that   the unit vector $u$ present in Section 2.1 
is just   $k/M$. {\em  This amounts to identify}   ${\widehat \del}$ of    (\ref{projec})  with  $\del _\perp$ of  2.1, and 
$     \hatx $  with  $  x_\perp  $. 

\noi Naturally we intent to remain as much as possible in the spirit of what is usually done in Newtonian mechanics,  so 
we aim at a  relativistic extension of several (if not all)  points of the list  a)-g). 
   We are not expecting to fulfill the whole list,  for example  reference to the vector-radius (point a) is not  a priori required,
 though welcomed when possible.

\medskip
\noi  For  simplicity   {\sl we assume from now on
that $V$ doesnot depend on $y \cdot P$.} It follows that   
\beq   V = V (\ztil ^2 , \ytil ^2 ,   \ztil  \cdot     \ytil ,       P^2  )            \label  {varunipot}    \eeq
 This restriction entails that  $G$  and  $\dron W / \dron p_0$ identically vanish; formula   (\ref{Tcdm}) reduces to
\beq   T(\lam ) =   \Ebr  \lam   +   {1 \over M}  \int    F d \lam  
\label{newTcdm}   \eeq

\bigskip

A clue toward a canonical  one-body formalism  is  provided by   the striking similarity of (\ref{evolztil})(\ref{evolytil})
 with    (\ref{evolxperp})(\ref{evolpperp}),  that  we  pointed out several decades ago \cite{cras}. 
 Actually they are the same formulas, up to notation.   The  Lie algebras generated by  $\ztil ^\alp  ,
 \ytil ^\beta  , P^ \gam $    under the Poisson brackets $ \{..,.. \} $ 
   and   by  $ x_\perp ^ \alp ,  p_\perp ^ \beta ,  M u^\gam$  under the brackets   $(..,..)$ are manifestly isomorphic,    
for instance compare  $\    \{ \ztil , \ytil \}  =  \widehat{\delta}$    with  
  $   ( x_\perp ^ \alp ,  p_\perp ^ \beta  ) =  \del _\perp  $. 
Since we have the same  Poisson bracket  structure, it is clear that from      (\ref{evolztil})(\ref{evolytil})
we obtain   (\ref{evolxperp})(\ref{evolpperp})   {\em  provided   we replace}
\beq     \lam  \mapsto  \tau ,  \qquad    \ztil \mapsto  x_\perp , \qquad \ytil \mapsto  p_\perp , \qquad    V \mapsto  W 
\label{presubstit}    \eeq
  where  
\beq  W = {\rm subs.}
 (V  |        \ztil = x_\perp ,  \ytil = p_\perp  , P^\alp = M u^\alp  )  
\label{substit}   \eeq
The substitution above introduces no dependence on  $x^0$ for  $W$, which is automatically constructed as stationary 
and  spherically symmetric.


\medskip
The  observation  presented  in   (\ref{presubstit})(\ref{substit})          makes the two-body problem equivalent to a one-body problem, 
at least {\em  in sofar as   {\em space}  relative variables  $\ztil,  \ytil$  are concerned.}

Let us try to go one step  further. 
 Since equations   (\ref{evolxperp})(\ref{evolpperp}) are just {\em  a subset} of the  system   (\ref{canon1})  generated by 
$H= \half p^2  +  W$,  it is natural to investigate  as  to  know to which extent  the 
{\em full binary dynamics}   can be  reduced  to  that of the  one-body  Hamiltonian system  described by 
{\em all} the  equations (\ref{canon1})   with  $W$  defined   in  (\ref{substit}).  To this end  we  introduce  
a  ficticious   pointlike body  moving in the external interaction   potential $W$.      
This  system,  governed by the  Hamiltonian generator $H$,  will be referred to as
 the  {\sl effective particle}.  {\sl Let $x_\eff ^\alp $ be its coordinates}.

\noi    Remark $\     $   
 W depends on $k^\alp$ as a parameter; to each   {\em binary  subsystem} characterized by fixing
 $P^\alp = k^\alp$  corresponds a distinct  one-body  dynamical system.    
       Formula   (\ref{substit})  uniquely   maps   
   a   {\em system of  equations}  (ruling the dynamical subsystem corresponding to a choice of $k^\alp$) 
to   another system of equations (deduced from $H$  and ruling the effective particle).
But their {\em solutions} are still to be specified by initial conditions (for instance fixing the numerical values of  first integrals). 
 As a result  
the   correspondance  between   pairs of  worldlines  and    one-body  motions  might  be affected  by some arbitrariness.
Nevertheless to each solution 
 $\ztil =\zeta (\lam, P^2 ),    \       \ytil   =\eta (\lam, P^2 ) \quad    $ of  (\ref{evolztil})(\ref{evolytil})    we can associate  
\beq  x_\perp  =  \zeta (\tau, M^2 ),    \qquad  \        p_\perp   =  \eta (\tau, M^2 )    \label{subssolspace}       \eeq 
which is a solution of    (\ref{evolxperp})(\ref{evolpperp}).

\medskip

It  will be  useful to distinguish, among two-body phase space functions,  those
that are  of the form  $J = {\cal J} (\ztil , \ytil , P )$. Let us call them  {\em functions of the special type}. 
As an example   $\ytil^2 ,  V,  N$  are  of the special type; in contrast $y \cdot P$ is {\em  not }.

\noi  We trivially extend to special-type functions the substitution (\ref{substit}) carried out in $V$, and write
\beq       {\rm subs.} J  =  {\cal J} (x_\perp , p_\perp , k )                \eeq

Let  $I =  {\cal I}  (\ztil , \ytil , P )     $    be    any  {\em first integral}  of the  special type.  Inserting  (\ref {solspace2}) into ${\cal I}$
we compute and find a result depending on the functions  $\zeta, \eta$  chosen  among possible solutions, but independent of $\lam$, say  $C$.
The one-body conterpart of $I$ is   $ {\rm subs. } I  =       {\cal I}  (x_\perp  , p_\perp  ,  k )$.
Inserting (\ref{subssolspace})  into    $ {\rm subs. } I  $  we automatically get a result independent of $\tau$ which is the same number $C$
 (see Appendix 1 for an example).  Thus
\beprop
When fixing the numerical value of a first integral {\em of the special type}, we assign the same value  to its one-body conterpart.
\enprop

\noi This remark can  be applied to $L^\alp$. Thus setting   $L^\alp   =  l^\alp$  implies  
$$ 
  \vareps  _ { \alp  \mu  \nu   \rho  }   k^\mu    x_\perp ^\nu  p_\perp ^\rho  =   l_\alp     $$
ensuring that the orbit of  the binary motion and that of   $x_\eff$  lay in the same two-dimensional plane.  

\noi Application to $N$ imposes  the same numerical values to $N$ and to  $p_\perp ^2  +  2  W$. 
But one-body Hamiltonian mechanics  yields  (\ref{m2E2}). Hence  
\beq  E^2  -  m^2  =  \Lam        \label{E2m2}         \eeq
where now $m$ is the mass of the effective particle, we shall refer to it as the {\em reduced mass}.
This formula obviously requires  
\beq  m^2  +  \Lam  > 0   \label{ineqm2}          \eeq


\bigskip

\subsubsection{ Motion  of the effective particle.}

\noi   As already noted previously, the {\em space part} of its equations of  motion is ruled by 
 (\ref{evolxperp}) (\ref{evolpperp})
analogous  to  (\ref{evolztil})(\ref{evolytil}).   Applying  substitution  (\ref{substit})   to  (\ref{solspace2})  we obtain  the  solutions
 \beq        x^i _\eff  =  \zeta ^i  (\tau , M ^2  )  ,    \qquad  \       p^i _\eff  =  \eta ^i  (\tau , M ^2  )       \label{xieff}   \eeq 
where  $\zeta ^i$  and   $\eta ^j $         are  the  functions defined in (\ref{solspace2}) as solutions of  (\ref{evolztil})(\ref{evolytil}).
The first formula  in    (\ref{xieff})  is  reminiscent of   (\ref{eobspace}) but  should not be confused with it,   because 
 $\tau$ {\em  may be different from} $\lam$.                

Now consider the  {\em time part} of the equations of motion.

\noi  On the one hand remember that our splitting of spacetime refers to the center-of-mass frame; thus we must identify
 the coordinate time of the  effective particle with the  time of the center of mass,  like  in
formula     (\ref{tcdmT}),  in other words
  \beq       x^0   _\eff =  x ^0 _\rel     =    T    \label{releff0}    \eeq
whith $T$  already  given by     (\ref{Tcdm}), where now   $G=0$  and  $F$ is a function of   the special type.

\noi On the other hand integrating the canonical equations of motion yields formula  (\ref{solx0}), where  now $\dron W  /  \dron  p_0$ vanishes thus
\beq   x^0 _\eff  =  E \tau  + {\rm const.}
\label{newsolx0}   \eeq

\noi  and  
$E$ is necessarily positive. 


\medskip

\noi  Comparing   (\ref{releff0})     with     (\ref{newsolx0})   and   taking  (\ref{newTcdm})  into account   yields  this   relation     
$  \disp  E  \tau      
  =   \Ebr \lam  +  {1 \over M } \int    
  F d\lam  +  {\rm const.}     $
This formula  cannot  yet  completely define $ \tau$ as a function of $\lam$ (or vice versa),  for two reasons:
on the one hand the  arbitrary integration constant involved in the  indefinite integral should be fixed; on the other hand  $E$ remains to be specified.

\noi First it is natural to demand, as a defining rule
 \beq    E  \tau        =   \Ebr \lam  +  {1 \over M } \int    _0  ^\lam    F d\lam    \label{condition}             \eeq 
 indeed  when $F$ is a constant of the motion  it allows trivially to identify $\tau$  with  $ \lam$.

\noi Then choosing $E$  in  terms of    the two-body constants of motion will    determine a  unique  map from  the  binary  pairs  of worldlines to 
 the  motions of  $x_\eff$,   or equivalently (by Theorem 1) a map  from  the  worldlines  of  $x_\rel$ to those of  $x_\eff$.  
We say that each  choice  of this kind  produces a {\bf version} of  the effective particle. 
   Naturally our freedom about  $E$  will be  constrained by obvious restrictions. For instance   
according to  (\ref{E2m2}), each choice of $E$ implies a unique  expression for  $m$ (and  conversely);
 it is clear that  this  $m$ must tend to $m_0$ in the nonrelativistic limit.

\noi  For practical  purpose we sometime prefer discussing the choice of $m$, which is the  relativistic reduced mass, and then derive the corresponding 
value of $E$.

\medskip
Now   a   relevant  question is  asking to which extent  the  {\em effective}  particle 
   can   coincide  with the {\em relative}  particle  defined by    (\ref{defeob}).  
Indeed (according to Theorem 1 above) in such a case the effective-particle motion  would  encode all information about 
the binary motion.

\noi  So let us compare  effective  and  relative  particles;  their time parts are equal, as seen in     (\ref{releff0}). But  $x^i _\eff $   satisfies
  (\ref{xieff})  whereas   $x^i _\rel $  is given by  (\ref{eobspace}).  Appearance of the same functions $\zeta $ ensures that   effective  and  relative  particles  have
 the  same  {\em orbit} in  the  vector space  orthogonal to $k^\alp$.  However the  {\em worldlines} are generally different due to equation  (\ref{condition})     which makes 
            $\tau$ generally  distinct  from  $\lam$,  implying that space and time coordinates are  differently related  in   $x_\eff$   and  in    $x_\rel$.  In other words the
 {\em  schedules} do not generally coincide; actually they coincide iff  $\tau = \lam$, which is possible by a suitable choice of $E$ provided that 
the right-hand side of    (\ref{condition}) is linear in $\lam$. 
 Let us summarize as follows
\betheo
Effective particle and relative particle have the same  orbit, but in general   they have different schedules.
\noi  They   have the same  worldline  iff   $\tau = \lam$, which is {\em not always} possible.
\entheo

Unfortunately,  requiring equality of  $\tau$ with $\lam$  is     a   very restrictive  condition. 
 It can be satisfied 
  for  all  motions  of the system  when    $F$ 
is a   first  integral,  provided we  choose
\beq    E =  \Ebr   +  {F \over M}                   \label{acadE}    \eeq

\noi  Alternatively it could be satisfied  without restrictions about $F$ but  {\em  only}  for
 {\em  circular} motions (this last point stems from the fact that $F$ remains constant on any circular orbit, as pointed out in proposition 5 of \cite{IJTP}).
An example is given in Appendix~2.

\noi
Having     $F$ constant {\em for all}  motions is  rather exceptional;    in particular  it is satisfied
 when     
  ${\dron V  /   \dron P^2}   =0   $.
In contrast circular motions exist under very large assumptions (see \cite{IJTP} ).    We can state     


\betheo   
In the academic case where   $F$  
is a  first integral of the binary  system,  the substitution  (\ref{substit}) together with identifications    $\tau =  \lam$  and   $E$  chosen  as in   (\ref{acadE}) above  
  makes the worldline derived from $H=\half  p^2 +    2W$ to coincide,  {\em for all motions of the system}, with   that of
 the  relative  particle defined in kinematical terms  through   (\ref{28ter})(\ref{tcdmT}).
\entheo 
However one must realize that in most realistic systems the assumptions of Theorem 3  {\em are not satisfied}. In the most  general situation we are left with
(\ref{condition}) which implicitly defines $\tau$ as a function of $\lam$ ( or vice versa)  but   this function can be complicated.


\noi Thus {\em in general it is not possible to demand that $x_\eff $ and $x_\rel$  coincide}.

\bigskip

 \noi    {\sl Equivalence}.

\noi  We saw  previously (Theorem 1)  
 how   the {\em relative} particle can be considered as  equivalent to the binary system.
 
\noi In order, for the {\em effective} particle, to deserve its name,   the question is whether binary motion can be similarly reduced to that of the {\em effective}  particle; 
  in other words:  is it possible to reconstruct the two-body  worldlines just by taking the effective particle motion (and center of mass) as input ?

\noi Since we are dealing with unipotential models we are already sure by Theorem~1   that  {\em  relative} motion encodes the worldlines of the two-body system.      
Thus in order to check  the reconstruction property of  {\em effective}  motion  it  is  sufficient to  observe  that 

\beprop  Once  the  effective particle's  version  has been choosen,  knowledge of   $x _\eff $  entails knowledge of  $x_\rel$.       \enprop

\noi Proof   $\      $    The choice of a version means  fixing   $E$ in terms of binary first integrals. Substitution (\ref{substit}) is manifestly invertible
so we  know the functions  $V$ and $W$ defined on their respective phase spaces.  Relative  and  effective  particles have the same orbit. All we still  need is a
one-to-one  correspondance between their  schedules; fortunately   formula  (\ref{condition})   yields  $\tau$ as a function of $\lam$
 as  well as (implicitly) the reverse.
  []

  \subsubsection{Choosing a version}

If we leave aside  the academic case  presented in  Theorem  3,  no choice of $E$ would make  effective particle   and  relative particle  identical.
Still we may   look for  a   "good  choice"   motivated by  some reasonable  requirement   or   by the 
  sake of simplicity; in any case  $m$ should  coincide with  $m_0$  in the nonrelativistic limit.
This  remark  will not  yet select  a unique 
version of the effective particle,  therefore it is convenient to sketch  a few possibilities (among others):

i)  $\     $The most simple choice seems to  consist  in   {\em defining} the reduced mass by the conventional  formula used in nonrelativistic mechanics, say 
$ m  =  m_0$.  According to    (\ref{E2m2}) and  (\ref{defLam}) this implies 

$$ E^2   =  m_0 ^ 2   +  { M^2 \over 4}   +   {  (m_1 ^2  -  m_2 ^2 )^2     \over  4  M^2}     -  \half (m_1 ^2  +  m_2 ^2 )    $$
Since $m_0 < m_1  \leq m_2 $ , for  negative $\Lam$  choosing   $m  =  m_0$ is submitted to the condition     $m_0 ^2   +  \Lam  \geq  0$,
 more restrictive than   (\ref{new48}).

ii) $\    $Alternatively we can adopt  the definition postulated by Todorov~\cite{todmass}      many years ago, $ \disp m =  m_T =  {m_1 m_2 \over  M} $.  In our notation  
(\ref{defmunu})  we can write   
  $\disp  m^2_T  =   {\mu ^2  - \nu ^2   \over  M^2} $.  Then   (\ref{E2m2})  and  (\ref{defLam})  entail
 $$  E^2  =  {1 \over  4 M^2} \     (4 \mu ^2 + M^4  - 4 \mu  M^2  ) =      {1 \over  4 M^2}  \              (M^2  - 2 \mu  )^2 $$     
Since $\disp   \dron  W \over \dron p_0 $ is zero, E cannot be negative, 
\beq    E  =    {1 \over  2 M}  \                 |M^2 - (m_1 ^2   +  m_2  ^2  ) |    \label{ETod}  \eeq     
which re-discovers  the expression found in \cite{todmass}  for the energy of the effective particle.   
It is well-known  that   $m_T \rightarrow   m_0$  in the nonrelativistic limit.

iii) $\    $  Another version, inspired by relativistic quantum mechanics,  may be considered  by demanding that, in case of bounded motion, the "unrest energy" 
of the effective particle, we mean $E -m$, be strictly equal to the binding energy of the binary system, namely $ M_{\rm bin}     =  M - (m_1 + m_2)$. 
Then we get 
$$ \disp   2m = {\Lam \over M_{\rm bin}}  -   M_{\rm bin}$$
which  implies  that   $ \Lam  -    M_{\rm bin}^2  $   must  have  the sign  of      $M_{\rm bin}$. 
Then  the development   (\ref{devlopM})   yields     
$$ {\Lam \over M_{\rm bin}  }   =2 m_0    +      O ( {\Lam \over  m_0^2})    $$
hence    $\disp  m = m_0  +   O ( {\Lam \over  m_0^2})$,  which reduces to $m_0$ in the nonrelativistic limit, as it  should.

iv)   We could  also   try choosing  $E$  such that $\tau = \lam$ on each circular orbit;  this can be actually  carried out for a toy model given in Appendix~2. 

\medskip


\section{Conclusion}

\noi

Relative motion is a  natural concept of  geometrical  origin, basically founded on  evolution of the radius-vector.
But  defining the relative {\em particle}  requires a  previous definition of center of mass.    Under this condition the relative particle is always  well-defined in geometrical terms, without specifying
in more details  how interaction is described  between the two bodies.  Assumptions A1 and A2, being   sufficiently general, are likely 	to accomodate a  large  number of  theories.
In principle, this (almost)  purely  kinematic  approach is  very general;  but  it would  remain academic  unless  we  address   the issue  of  {\em equivalence}.     
Considering   this question  in   the  general  framework of  predictive relativistic dynamics,  we realized  that   the  map  of the binary motion onto that of a single particle is
 not  always  unambiguously  invertible:  
some dynamical information (the  function $U$)  is needed in order to reconstruct the worldlines of the binary object  from  the motion of the relative particle.  

\noi  This limitation led us to focus on unipotential models of  mutual interaction, and   in this  context  it was indeed possible  to derive some nice properties  of the relative particle
 (Theorem~1 and Proposition~3) satisfying the points c) and e) of the list in  Section 3.
However  some  features  of  the  classical    theory, namely the reduced mass and  the canonical  formalism,  had  no   relativistic  conterpart at this stage.

    This  shortcoming  was   calling  for    an alternative approach; therefore we  proposed  that beside the relative particle one  considers  another ficticious body, referred to as 
{\em effective}, which is,  by a simple rule,   constructed  as a  Hamiltonian system.    This procedure  involves some   arbitrariness  related to the ajustment of a constant of the motion,   
 which can be interpreted as   some    freedom in the choice  of the relativistic reduced mass.   One of the possible choices retrieves Todorov's effective particle originated from QED  \cite{todmass} and  
further  given a worldline content \cite{tod} in the framework of relativistic constraint dynamics.

The relative particle is   unambiguously defined whereas  the effective particle is a priori affected by an arbitrariness, the solution of which we had to discuss. 
Fortunately it  turns out that relative and effective particle have anyway   the  {\em same orbit}  and differ only by their schedules; only in particular cases they completely coincide.
So in practice various   versions of  the  effective particle  are  more or less equally useful;  further investigation  might bring out a preferrence  among the possible choices  just  listed  above 
(a list which is not exhaustive).
 
Most part of our  picture has been elaborated in  the context  of  rather simple  hypotheses;  more investigation  is needed  for instance   if  we  relax the assumption that  
${\dron V  / \dron  ( y \cdot P)}$ vanishes.  But we hope that  the present work will  already clarify  several   ideas about relative
 motion in  Minkowski spacetime.

\medskip
$$  \qquad    $$
{\bf  Appendix   1}

\noi Consider the toy model presented in \cite{IJTP},  say
\beq V = \chi   \   \sqrt{ P^2}  \ztil ^2       \label{oscar}      \eeq
with $\chi$ a positive string  constant.
Fixing  $P^\alp  =  k^\alp$ and fixing  $L^\alp = l^\alp$ orthogonal to it,  the solution to  (\ref{evolztil})(\ref{evolytil})  is
\beq
\ztil =  \zeta (\lam)  =   A  \   \sin (\Ome \lam + \Phi) +  B  \   \cos (\Ome \lam + \Phi)
\label{ztiloscar}   \eeq
\beq
\ytil =  \eta (\lam ) =    A  \Ome  \    \cos (\Ome \lam + \Phi)  - B  \Ome \    \sin (\Ome \lam + \Phi)
\label{ytiloscar}
\eeq                                             
The orbital plane is orthogonal to both $k$ and $l$,  and 
 $A,    B$ are  mutually orthogonal spacelike constant vectors in that plane   ( $|A|$ and $|B|$ are  the half-axes of an ellipse).   $\Phi$ is a scalar constant,
 moreover we have  $\        \Ome = \sqrt{2 \chi  M } \    $.      
 Consider the first integral
$N =  {\cal N}   \   (\ztil , \ytil , P )  =       \ytil ^2 + 2V$.
From   (\ref{ztiloscar})(\ref{ytiloscar})  we compute   
$\ztil ^2  $ and $\ytil ^2$,  hence we find  that  the numerical value of $N$, say  
$$ <N > =   2 \chi M (A^2  + B^2  )   $$
is (as expected) independent of  $\lam$.

\noi  In the substitution     (\ref{presubstit})(\ref{substit})   the one-body conterpart of  $V$ is 
$W =   \chi   \      M  x_\perp ^2$  and the conterpart of  $N $ is
\beq  {\rm subs.}  N  =   {\cal N}   \   (x_\perp ,  p_\perp , k )  =   p_\perp ^2  +   2 \chi  M   x_\perp ^2    \label{subsN} \eeq
To the solution (\ref{ztiloscar})(\ref{ytiloscar}) of the binary problem we associate 

\beq     x_\perp =  \zeta (\tau)  =   A  \   \sin (\Ome \tau + \Phi) +  B  \   \cos (\Ome \tau + \Phi)
\label{xoscar}   \eeq
\beq
p_\perp =  \eta (\tau ) =    A  \Ome  \    \cos (\Ome \tau + \Phi)  - B  \Ome \    \sin (\Ome \tau + \Phi)
\label{poscar}
\eeq
Inserting    (\ref{xoscar})(\ref{poscar})  into (\ref{subsN}) yields of course 
  $$  p_\perp ^2  +   2 \chi  M   x_\perp ^2   =   2 \chi M (A^2  +B^2  )  $$
manifestly independent of $\tau$  and  identical to $<N>$.
 
\medskip
\noi   {\bf  Appendix~2}

\noi
For the same model,  beside $\Lam$ we have another first integral  $j^2$ given by  (\ref{defj2}), and  
\beq  \half (A^2 +  B^2 ) = - { \Lam   \over  4 \chi M} =  { <N>   \over  4 \chi M}  ,      \qquad  \qquad        
 A^2  B^2  =   {j^2  \over  \Ome ^2}   \label{Lamjax}      \eeq
Straightforward calculations  yield (\cite{IJTP})
$$  F = \chi M  \left(  A^2 \sin ^2  (\Ome \lam +  \Phi ) + B^2  \cos ^2  (\Ome \lam +  \Phi )  \right)     $$
a primitive of this function is
$$  \int _ {0} ^\lam   F  d \lam   =   \chi M  \left( {A^2 +  B^2   \over 2} \lam   +   
 {B^2 -  A^2   \over  4  \Ome}   \sin   (2 \Ome \lam  + 2 \Phi ) + { A^2  -  B^2  \over  4  \Ome}  \sin  (2 \Phi )    \right)  $$

\noi  Circular orbits are characterized by  $A^2  = B^2$, so if we define
$ F_{\rm cir}   =  {N \over 4}$, this quantity, defined  {\em  on  the whole phase space},  is a first integral which coincides with $F$
 on any circular orbit.

\noi 
Thus if we choose (for all orbits) 
$$ E = \Ebr +  {  F_{\rm cir} \over  M}  =     \Ebr +  {N \over  4M}      $$
equation  (\ref{condition})  allows for  having  that $\tau = \lam$ on any circular orbit.





\end{document}